\documentclass[prd,preprint,a4paper,superscriptaddress,nofootinbib,showpacs,amsfonts,amssymb, 11pt]{revtex4}

\usepackage{url}
\usepackage{graphicx}
\def\fs{\footnotesize}

\begin{document}

\title{\Large Accelerated Cosmological Models  in \\ Ricci squared Gravity}

\date{\today}

\author{Gianluca ALLEMANDI}
\email{allemandi@dm.unito.it}
\affiliation{{\fs Dipartimento di Matematica},
{\fs Universit\`a di Torino}\\{\fs Via C. Alberto 10, 10123
TORINO (Italy)}}
\author{Andrzej BOROWIEC}
\email{borow@ift.uni.wroc.pl}
\affiliation{\fs Institute of Theoretical Physics, University of Wroc{\l}aw\\
Pl. Maksa Borna 9, 50-204  WROC{\L}AW (Poland)}
\author{Mauro FRANCAVIGLIA}
\email{francaviglia@dm.unito.it}
\affiliation{{\fs Dipartimento di Matematica},
{\fs Universit\`a di Torino}\\{\fs Via C. Alberto 10, 10123
TORINO (Italy)}}

\pacs{98.80.Jk, 04.20.-q}

\begin{abstract}
Alternative gravitational theories described by
 Lagrangians depending on general functions of the Ricci scalar
have been proven to give coherent theoretical models to describe the
experimental evidence of  the acceleration of universe at present
time. In this paper we proceed further in this analysis of
cosmological applications of alternative gravitational theories
depending on (other) curvature invariants. We introduce Ricci
squared Lagrangians in minimal interaction with matter (perfect
fluid); we find modified Einstein equations and consequently
modified Friedmann equations in the Palatini formalism. It is
striking that both Ricci scalar and Ricci squared theories are
described in the same mathematical framework and both the
generalized Einstein equations and generalized Friedmann equations
have the same structure. In the framework of the cosmological
principle, without the introduction of exotic forms of dark energy, we thus obtain
modified equations providing values of $w_{ eff}<-1$ in accordance
with the experimental data. The spacetime  bi-metric structure plays
a fundamental role in the physical interpretation of  results and
gives them a clear and very rich geometrical interpretation.

\end{abstract}

\maketitle

\section{Introduction}
 In this paper we try to better understand and to analyze  alternative theories of gravity depending on
 higher-order terms in the curvature invariant $R^{(\mu \nu)} R_{(\mu \nu)}$, in relation with some very
 interesting and  possible cosmological application and, in particular, in relation with their capability
 to explain the cosmological acceleration of the universe, both in early times (inflation) and in present
 time universes. Nevertheless we will focus our attention on the possible theoretical explanation of the
 present cosmological acceleration. \\
Recent astronomical observations have shown that the universe is
accelerating at present time (see \cite{Perlmu} and  \cite{Riess}
for supernova  observation results;  see \cite{Spergel} for the
observations about the anisotropy spectrum of the cosmic microwave
background (CMBR); see \cite{Verde} for  the results about the power
spectrum of large-scale structure).  Physicists have  thus to face
the evidence of the acceleration of the universe and should  give a
coherent theoretical explanation to these experimental results: a
problem which  up to now seems to be still unsolved! General Relativity in
interaction with a perfect-fluid like matter and the cosmological
principle, providing the standard cosmological models, fail  to give
by their own a theoretical framework  to explain the  acceleration
of the universe. We are thus forced to introduce some kind of
\textit{dark matter or dark energy}, which are responsible for the
acceleration of the universe, or to modify
General Relativity such that acceleration is predicted (see for example \cite{Carrol1}). \\
Dark matter or dark energy models have been deeply investigated in
relation with their capability of explaining the acceleration of the
universe (see \cite{darkmatter} and references therein), however up
to now there are no satisfactory experimental evidences of the
presence of the predicted amount of dark energy in the universe. The
real nature of dark energy, which is required by General Relativity
in this cosmological context,  is unknown but it is fairly well
accepted that dark energy should behave like a fluid with a large
negative pressure. The dark energy models with effective equation of
state $w_{eff}$ (which determines the relation between pressure $p$
and density of matter $\rho$) smaller than $w_{eff}<-1$ are
currently preferrable, owing to the experimental results of  \cite{Spergel}. \\
On the other side the simplest way of obtaining accelerated expansions within General Relativity is to introduce a
positive cosmological constant \cite{lambda}, an introduction which leads however to some theoretical and experimental
problems and contradictions (see for example \cite{Carrol1} and \cite{lambda}). We just want to stress here that models
with a constant cosmological constant are not able to explain the evolution between different epochs of the universe,
characterized by different values of acceleration (deceleration).  \\
The other possibility is to assume that we do not yet understand
gravity at large scales, which means that General Relativity should
be modified or replaced by alternative gravitational theories of
Gravity when the curvature of spacetime is small (see for example
\cite{staro}, \cite{Nojiri}, \cite{brane} and references therein),
providing modified Friedmann equations. Hints in this direction are
suggested moreover from the quantization on curved spacetimes, when
interactions among the quantum fields and the background geometry or
the self interaction of the gravitational field are considered. It
follows that the standard Hilbert-Einstein Lagrangian has to be
suitably modified by means of corrective terms, which are essential
in order to remove divergences \cite{staro}. These corrective terms
result to be higher-order terms in the curvature invariants such as
$R$, $R^{\mu \nu} R_{\mu \nu}$, $R^{\mu \nu \alpha \beta} R_{\mu \nu
\alpha \beta}$, $R\square^lR $, or non minimally coupled terms
between scalar fields and the gravitational field. It is moreover
interesting that such corrective terms to the standard
Hilbert-Einstein Lagrangian can be predicted in higher dimensions by
some time-dependent compactification in string/M-theory (see
\cite{Nojiri}) and corrective terms of this type
arise surely in brane-world models with large spatial extra
dimensions \cite{brane}. As a matter of facts, if these brane models
are the low energy limit of string theory, it is likely that the
field equations include in particular the Gauss-Bonnet term, which
in five dimensions is the only non-linear term in the curvature
which yields second order field equations. In this framework
Gauss-Bonnet corrections should be taken into account and
cosmological models deriving from the Gauss-Bonnet have been recently studied; see \cite{gauss} and
references therein. \\
As an alternative to extra dimensions, it is also possible to
explain the modification to  Friedmann equations (which could
provide a theoretical explanation for the acceleration of the
universe) by means of a modified theory of four dimensional gravity.
The first attempts in this direction were performed by adding to the
standard Hilbert-Einstein Lagrangian analytical terms in the Ricci
scalar curvature invariant  \cite{carrol2}. A simple task to modify
General Relativity, when the curvature is very small, is hence to
add to the Lagrangian of the theory a piece which is proportional to
the inverse of the scalar curvature $1 \over R$ or to replace the
standard Hilbert-Einstein action by means of polynomial-like
Lagrangians, containing both positive and negative powers of the
Ricci scalar $R$ and logarithmic-like terms. Such theories have been
analyzed and  studied both in the metric \cite{metricfR} and the
Palatini formalisms \cite{ABFR}, \cite{palatinifR}. It results that
both in the metric and the Palatini formalism they provide a
possible theoretical explanation to the present time acceleration of
the universe. Moreover a mechanism ruling the present dark energy
dominance\footnote{Taking into account the transition of the universe from a decelerated era to an accelerated era, a scenario
 with $w_{eff}$ transitting from values below $-1$ to values above $-1$ is actually preferable \cite{xinmin}.} 
(due to the universe expansion) and the present cosmic
acceleration has been proposed in this framework; see  \cite{darkdpm}.\\
A discussion is open on the physical reliability of the Palatini
and/or the metric formalism \cite{Flanagan} and on the physical
relevant frame both in the metric and the Palatini formalism
\cite{Magnano}. Up to now it appears that  the aforementioned metric
approach\footnote{We remark that field equations are in that case
fourth order field equations.} leads to results which are in
contrast with the solar system experiments  and also that the
relevant fourth order field equations suffer serious instability
problems \cite{metricfR}. On the contrary the Palatini formalism
produces second order field equations which are not afflicted by
instability problems and are in acceptable accordance with the
results of the solar system experiments \cite{palatinifR}.  A
discussion is actually open on the accordance of the Palatini
formalism with the electron-electron scattering experiments
\cite{Flanagan}.\\
\\
The importance of modified theories of gravity depending on general
analytical functions of the Ricci scalar is also related with the
possibility of avoiding singularities in these cosmological models
\cite{kerner} and in the interpretation of  black holes entropy in
this context \cite{BNOV}.
Recently, an explanation of the present day acceleration of the universe has been 
moreover formulated in the framework of non-symmetric gravitational theories \cite{moffat} 
and in modified theories depending on the determinant of the Ricci tensor \cite{dolgov}.\\
\\
\noindent Encouraged by  recent developments of cosmological
applications of alternative theories of Gravity we consider in this
paper Ricci squared Lagrangians in minimal interaction with matter,
which have been deeply analyzed in \cite{BFFV} in the vacuum case.
As we already said before such Lagrangians are deeply related with
quantum field theory: to remove divergences one has to add
counterterms to the Lagrangian which depend not only on the Ricci
scalar but also on the Ricci and the Riemann tensors  \cite{staro}.
It was proven in \cite{BFFV} that Ricci squared Lagrangians  provide
second order field equations in the Palatini formalism, such that
the universality of Einstein equations and the universality of the
Komar energy-momentum complex hold in vacuum.  These remarkable
results have important implications also in cosmological models. They
imply that, in some sense, field equations for Ricci squared
Lagrangians reproduce (apart conformal transformations) the standard
Einstein field equations in the vacuum case, while in the presence
of matter this equivalence might be  broken. The geometrical
structure of the spacetime manifold is very rich and it is endowed
with an anti-K\"ahlerian structure, deriving directly from the
variational principle of Ricci squared Lagrangians (see
\cite{antiK}). Spacetime turns out to have a bi-metric structure, or
better a so-called \textit{metric compatible almost-product} as well
as an \textit{almost-complex structure with a Norden metric}.  The
geometrical structure of spacetime is moreover characterized by a
scalar-valued structural equation, which is simply obtained by
contracting field equations with the metric \cite{BFFV} and  controls the solutions of 
field equations. Lagrangians based on higher-order Ricci scalars which led to higher-order
metric compatible polynomial structures have been considered both in purely metric and Palatini
formalism in \cite{BFS}.\\
It was moreover shown in \cite{antiK} and \cite{cosmolla} that Ricci
squared theories in vacuum give field equations equivalent to
Einstein field equations with a cosmological constant, the value of
which is fixed once the structural equations are solved and one
particular solution of the structural equations is chosen. This fact
is no longer true in the case of interaction with generic matter,
where the solutions of the structural equation are dynamical (the
same happens in the case of non-linear Lagrangians in the Ricci
scalar; see \cite{ABFR} and references therein).  The equivalence
with Einstein field equations is hence broken and we obtain modified
field equations, depending on the stress-energy tensor of matter
involved in the theory. Nevertheless, we remark that field equations
are once again second order field equations in the metric field. \\
For cosmological applications we consider the physical metric $g$ to
be a Robertson-Walker metric and the stress-energy tensor of matter
to be a perfect fluid one. In this particular framework deriving
from the cosmological principle we obtain that the Levi-Civita
metric $h$ is  \textit{conformal} to the physical metric $g$, apart
for a rescaling factor of the cosmological time. From our
construction it follows however that the signature of $h$ can be
arbitrarily chosen (it can be either  Riemannian, or Lorentzian or
Kleinian, apart from some restrictions deriving from field equations).
We are consequently able to introduce a generalized Hubble constant and modified Friedmann equations.
A comparison with the $f(R)$ theories is immediate. It is striking to notice that modified Friedmann
equations are once more first order field equations, which prevent the appearing of instabilities as it
has already been shown in the case of Ricci scalar theories in \cite{palatinifR}. This is an important
consideration giving the Palatini formalism a deeper physical significance, in view of cosmological applications.\\
An explicit example dealing with power Lagrangians  in the Ricci squared invariant $f(S)=\beta S^n$ is analyzed in detail and the Hubble constant is derived. It results that the deceleration parameter can be negative if particular values of $n$ are chosen. Moreover we obtain values of $w_{eff}$ that can be suitably fitted to the experimental results of \cite{Perlmu}. Considerations are exposed about the  \textit{frame changing}, which means choosing $h$ to be a FRW metric instead of $g$. Field equations and cosmological parameters  are obtained and discussed also in that alternative (\textit{Jordan}) frame.\\
\\
The paper is organized as follows: we start in section 2 by considering the case of $f(R)$ Lagrangians in a new matrix formalism, with the introduction of an operator $P$ modifying the Einstein field equations \cite{ABFR}. We pass in section 3 to the more complicated case of Ricci squared $f(S)$ Lagrangians and we analyze the field equations and the structure of spacetime in the case of interaction with matter. We proceed in section 4 with cosmological applications and we obtain modified Friedmann equations.  In section 5  we discuss the relevant example of polynomial Lagrangians in the Ricci squared invariant. In section 6 we consider the theory in the alternative Jordan frame, where $h$ is assumed to be a priori the FRW physical metric.


\section{Cosmological models in $f(R)$ Gravity}
 We start considering non-linear Lagrangians in the Ricci scalar invariant $f(R)$, already treated and developed in \cite{ABFR} and in \cite{FFV} in the vacuum case.
We think that it is worth summarizing those theories in order to have a comparison with Ricci squared theories here developed and analyzed in detail. We moreover modify the formalism introduced in \cite{ABFR} to treat both Ricci  scalar and Ricci squared theories in the same mathematical framework. \\
The action for $f(R)$ Gravity is introduced to be:
\begin{equation} \label{lagrfR}
A=A_{\mathrm{grav}}+A_{\mathrm{mat}}=\int (\sqrt{\det g}f(R)+2\kappa L_{\mathrm{mat}}(\Psi))d^{4}x
\end{equation}
where $R\equiv R( g,\Gamma) =g^{\alpha\beta}R_{\alpha \beta}(\Gamma )$ is the \textit{generalized Ricci scalar} and $
R_{\mu \nu }(\Gamma )$ is the Ricci tensor of a torsionless connection $\Gamma$.
The gravitational part of the Lagrangian is controlled
by a given real analytic function of one real variable $f(R)$,
while $\sqrt g$ denotes the scalar density $\mid\det\parallel g_{\mu\nu}\parallel\mid^{\frac{1}{2}}$ of weight $1$.
The total Lagrangian contains also a matter part $L_{\mathrm{mat}}$ in minimal interaction with the gravitational field, depending on matter fields $\Psi$ together with their first derivatives and  equipped with a gravitational coupling constant $\kappa=8\pi G$.\\
Equations of motion, ensuing from the first order \'a la Palatini formalism
are (we assume the  spacetime manifold to be a Lorentzian manifold $M$ with $\dim M=4$; see \cite{FFV}):
\begin{equation}
f^{\prime }(R)R_{(\mu\nu)}(\Gamma)-\frac{1}{2}f(R)g_{\mu \nu
}=\kappa T_{\mu \nu }  \label{ffv1}
\end{equation}
\begin{equation}
\nabla _{\alpha }^{\Gamma }(\sqrt{\det g}f^{\prime}(R)g^{\mu \nu })=0
\label{ffv2}
\end{equation}
where $T_{\mu\nu}=-\frac{2}{\sqrt g}\frac{\delta L_{\mathrm{mat}}}{\delta g_{\mu\nu}}$
denotes the matter source stress-energy tensor and $\nabla^{\Gamma}$ means covariant derivative with respect to $\Gamma$. \\
We shall use the standard notation
denoting by $R_{(\mu\nu)}$ the symmetric part of $R_{\mu\nu}$, i.e.
$R_{(\mu\nu)}\equiv \frac{1}{2}(R_{\mu\nu}+R_{\nu\mu)})$.
In order to get (\ref{ffv2}) one has to additionally assume
that $L_{\mathrm{mat}}$ is functionally independent of $\Gamma$; however it may
contain metric  covariant derivatives $\nabla^g$ of fields.
This means that the matter stress-energy tensor $T_{\mu\nu}=T_{\mu\nu}(g,\Psi)$
depends on the metric $g$ and some matter fields denoted
here by $\Psi$, together with their derivatives.
From (\ref{ffv2}) one sees that $\sqrt{\det g}f^{\prime }(R)g^{\mu \nu }$
is a symmetric twice contravariant tensor density of weight $1$, so that if
not degenerate one can use it to define a metric $h_{\mu \nu}$ such that
the following holds true
\begin{equation}\label{h_met}
\sqrt{\det g}f^{\prime }(R)g^{\mu \nu}=\sqrt{\det h}h^{\mu \nu }
\end{equation}
This means that both metrics $h$ and $g$ are conformally equivalent. The
corresponding conformal factor can be easily found to be $f^{\prime}(R)$
(in $\dim M=4)$ and the conformal transformation results to be:
\begin{equation}\label{h_met1}
h_{\mu \nu }=f^{\prime}(R)g_{\mu \nu }
\end{equation}
Therefore, as it is well known, equation (\ref{ffv2}) implies that  $\Gamma =\Gamma _{LC}(h)$ and $R_{(\mu\nu)}(\Gamma)=R_{\mu \nu }(h)\equiv R_{\mu\nu}$.
Let us now introduce a (1,1)-tensorfield  $P$ by
\begin{equation}\label{P_mat}
P_{\nu }^{\mu }=g^{\mu \alpha }R_{\alpha \nu }( h )
\end{equation}
so that (\ref{ffv1}) re-writes as
\begin{equation}\label{P_mat2}
f^{\prime }(R)P_{\mu }^{\nu }-\frac{1}{2}f(R)\delta _{\mu }^{\nu }=\kappa
\widehat{T}_{\mu }^{\nu }
\end{equation}
where, with an abuse of notation, $\hat T=\hat{T}_{\mu }^{\nu }=g^{\mu \alpha }T_{\alpha \nu }$\ \ and  from (\ref{P_mat}) we obtain that $R=\mathrm{tr}P$. \\
Equation (\ref{P_mat2}) can be supplemented by the scalar-valued equation
obtained by taking the trace of (\ref{P_mat2}),
(we define $\tau=\mathrm{tr}\hat T$)
\begin{equation} \label{structR}
f^{\prime }(R)R-2f(R)=\kappa g^{\alpha\beta}T_{\alpha\beta}\equiv \kappa\tau
\end{equation}
which controls solutions of (\ref{P_mat2}). We shall refer to this scalar-valued equation as the \textit{structural equation} of spacetime. The structural equation (\ref{h_met1}), if explicitly solvable, provides an expression of $R=F(\tau)$ and consequently both $f(R)$ and $f^\prime (R)$ can be expressed in terms of $\tau$.
More precisely, for any real solution $R=F(\tau)$ of (\ref{structR}) one has that the operator $P$ can obtained from the matrix equation (\ref{P_mat2}):
\begin{equation}\label{alg1}
P =\frac{f(F(\tau))}{2f^{\prime}(F(\tau))}
 I+\frac{\kappa }{f^{\prime}(F(\tau))}\hat T \end{equation}
Now we are in position to introduce the generalized Einstein  equations under the  form
\begin{equation}\label{gen_Ein}
g_{\mu \alpha }P_{\nu }^{\alpha }=R_{\mu \nu }\left( h\right)
\end{equation}
where $h_{\mu\nu}$ is given by (\ref{h_met1}) and $P_{\nu }^{\mu }$ is obtained  from the algebraic equations (\ref{structR}) and (\ref{alg1}) (for a given $g_{\mu \nu }$ and $T_{\mu\nu}$); see also \cite{ABFR} and \cite{FFV}. For the matter-free case we  find
that $R=F(0)$ becomes a constant   implying  that the two metrics are proportional and the operator $P$ is proportional to the Kronecker delta. Equation (\ref{gen_Ein}) is  hence nothing but Einstein equation for the metric $g$,
almost independently on the choice of the function $f(R)$, as already obtained in \cite{FFV}.
Also the standard Einstein  equation with a cosmological constant $\Lambda$ can be recasted  into the form (\ref{gen_Ein}). It corresponds to the choice $f(R)=R-\Lambda$. These properties justify  the name of generalized Einstein equation given to (\ref{gen_Ein}). In the presence of matter equation  (\ref{gen_Ein}) expresses a deviation for the metric $g$ to be an Einstein metric as it was discussed in \cite{ABFR}. It can be otherwise interpreted as an Einstein equation with additional stress-energy contributions deriving from the modified gravitational Lagrangian \cite{palatinifR}, or possibly as a modified theory of gravity with a time dependent cosmological constant.


\subsection{Cosmological applications of first-order non-linear gravity \label{cossec}}
We give here a brief summary of the results obtained in \cite{ABFR}, where we refer the reader for further details.  We assume  $g$ to be  the Friedmann-Robertson-Walker (FRW) metric which (in spherical coordinates) takes the standard form:
\begin{equation}
g=-d t^2+a^2 (t) \Big[ {1 \over {1-K r^2}} d r^2+ r^2 \Big( d \theta^2 +\sin^2 (\theta) d \varphi^2  \Big) \Big] \label{RW1R}
\end{equation}
where $a (t)$ is the so-called \textit{scale factor} and $K$ is the space curvature ($K=0,1,-1$).
We further choose  a perfect fluid stress-energy tensor for matter:
\begin{equation} \label{Tmunuuu}
T_{\mu\nu}=
(\rho+p)u_\mu u_\nu+pg_{\mu\nu}
\end{equation}
where $p$ is the pressure, $\rho $ is the density of matter and $u^\mu$ is a co-moving fluid vector, which in a co-moving frame
($u^\mu=(1,0,0,0)$) becomes simply:
\begin{equation}
T_{\mu \nu}=
\left(
\begin{array}{clcr}
\rho&0&0&0\\
0&\frac{pa^2 (t)}{1-K r^2}&0&0\\
0&0&pa^2 (t)r^2&0\\
0&0&0&pa^2 (t)r^2\sin^2 (\theta)
\end{array}
\right) \label{TmunuR}
\end{equation}
The metric $h$ turns out to be conformal to the FRW metric $g$ by means of the conformal factor $f^\prime (R)$, which
can be moreover expressed in terms  of $\tau $ by means of (\ref{structR}) and finally as a function of time $b(t)=f^\prime (R(\tau))$, by an abuse of notations. From (\ref{gen_Ein}) we can obtain
 an analogue of the Friedmann equation under the form
\begin{equation}
\hat{H}=\Big(\frac{\dot a}{a} +\frac{\dot b}{2b}\Big)^2=
\frac{\kappa}{3b}\Big[\rho+\frac{f(\tau)+\kappa\tau}{2\kappa}\Big]- \frac{K}{a^2}  \label{MHC}
\end{equation}
which can be seen as a generalized definition of a \textit{modified Hubble constant} $\hat{H}=\Big(\frac{\dot a}{a} +\frac{\dot b}{2b}\Big)$,
taking into account the presence of the conformal factor $b(t)$ which enters into the
definition of the conformal metric $h$  (see \cite{ABFR} for details). This equation reproduces, as expected, the standard Einstein equations in the case $f(R)=R$.\\
Considering the particular example  $f(R) =\beta R^n$ we have obtained that the Hubble constant for the metric $g$ can be locally calculated to be:
\begin{eqnarray}
H^2=\varepsilon r(n,w) a^{\frac{-3(w+1)}{n}}
-s(n,w) \frac{K}{a^2}  \label{MFRn}
\end{eqnarray}
where:
$$
\cases{
r(n,w)=\frac{2n}{3(3w-1)[3w(n-1)+(n-3)]}
\left[  \frac{-\kappa(3w-1)}{\beta (2-n)} \right]^{\frac{1}{n}} \cr
s(n,w)= \left[ \frac{2n}{3w(n-1)+(n-3)} \right]^2
}
$$
are functions of the power $n$ and of the equation state of matter, through $w$.
We remark  that $\varepsilon={\rm sign} R = 1$ for odd values of $n$ and, on
the contrary, $\varepsilon=\pm 1$ for even values of $n$; see \cite{ABFR}  for details.
The deceleration parameter can be obtained from the Hubble
constant by means of the following relation:
\begin{equation}
q(t):=- \left(1+\frac{\dot{H}(t)}{H^2(t)} \right)=-\left(
\frac{\ddot{a}(t)}{a(t) H^2(t)} \right) \label{qhub}
\end{equation}
and from (\ref{MFRn}) it turns out to be formally equal to:
\begin{equation} \label{decpowR}
q(t,w,n) =-1+ \frac{\frac{3(1+w)}{2n}\epsilon r(n,w)  a^{-\frac{3(1+w)}{n}}-s(n,w){K}a^{-2} }{\epsilon r(n,w)  a^{-\frac{3(1+w)}{n}}-  s(n,w){K} a^{-2}}
\end{equation}
It follows that when the $a^{-2}$ term dominates over $a^{-\frac{3(1+w)}{2n}}$ the deceleration parameter 
results to be positive, i.e. $q(t,w,n)\rightarrow 0$. On the contrary,   when the term $a^{-\frac{3(1+w)}{n}}$ dominates over $a^{-2}$ (or in the case $K=0$ corresponding to spatially flat spacetime) the deceleration parameter results to be:
\begin{equation} \label{decpowlimR}
q(w,n) =-1+ \frac{3(1+w)}{2n} \label{qqqquti}
\end{equation}
which is  negative for $n<0$ or $n>\frac{3(1+w)}{2}>0$ owing to the positivity of $(1+w)$ for 
standard matter; see \cite{ABFR}. This implies that the accelerated behavior of the universe is predicted 
in a suitable limit. In particular it follows that super-acceleration ($q<-1$) can be achieved only for 
$n<0$. The effective $w_{eff}$ can be obtained (as in \cite{carrol2}) by means of simple 
calculations from (\ref{MFRn}) and  (\ref{qqqquti}). It results to be, for this theory:
\begin{equation}
 w_{eff}={2\over 3} q(n, w)-{1 \over 3}=-1+{(w+1) \over n}\label{weffR}
\end{equation}
We remark that the range of $-1.45<w_{eff}<-0.74$ for dark energy, stated in \cite{Spergel}, can be
easily recovered in this theory by choosing suitable and admissible values\footnote{As already explained in \cite{ABFR} the parameter $n$ should not be an integer, it can be any real number satisfying some reliability conditions; see \cite{ABFR} for further discussions and details.} of $n$. We refer to \cite{ABFR} for physical considerations and  for more detailed discussions and examples concerning polynomial-like Lagrangians in the generalized Ricci scalar. \\


\section{Ricci squared Lagrangians in minimal interaction with matter fields \label{R2bemetr}}

We consider now the action functional:
\begin{equation}
A=A_{\mathrm{grav}}+A_{\mathrm{mat}}=\int (\sqrt{\det g}f(S)+2\kappa L_{\mathrm{mat}} (\Psi) )d^{4}x
\end{equation}
where $S\equiv S\left( g,\Gamma \right) =g^{\mu \alpha }R_{\left( \alpha \nu
\right) }(\Gamma )g^{\nu \beta }R_{\left( \beta \mu \right) }(\Gamma )$ and $
R_{\mu \nu }(\Gamma )$ is, as above, the Ricci tensor of a torsionless connection $\Gamma $ (see discussion after formula  (\ref{lagrfR})).
The gravitational part of the Lagrangian is controlled
by a given real function of one real variable $f(S)$; see \cite{BFFV}.
Under the same assumptions of \cite{BFFV} and in $4$-dimensional spacetimes $M$ ($\dim M=4$) equations of motion ensuing from the variational principle in the Palatini formalism are  \cite{BFFV}:
\begin{equation}
2f^{\prime }(S)g^{\alpha \beta }R_{( \mu \alpha)}(\Gamma )R_{(\beta\nu)}(\Gamma)-
\frac{1}{2}f(S)g_{\mu \nu }=\kappa T_{\mu \nu }  \label{bffv1}
\end{equation}
\begin{equation}
\nabla _{\sigma }^{\Gamma }(\sqrt{\det g}f^{\prime }(S)g^{\mu \alpha
}R_{(\alpha \beta)}(\Gamma)g^{\beta\nu })=0  \label{bffv2}
\end{equation}
where $T_{\mu\nu}=-\frac{2}{\sqrt g}\frac{\delta L_{\mathrm{mat}}}{\delta g_{\mu\nu}}$
denotes again the matter source stress-energy tensor.
The above system of equations splits, as before,
into an algebraic part (\ref{bffv1}) and a differential one (\ref{bffv2})
for the unknown variables $g$ (the metric)  and $\Gamma$ (the connection).\\
Following the general strategy elaborated for the matter-free case  \cite{BFFV} and \cite{FFV}
(see also \cite{ABFR}) let us notice that $\sqrt{\det g}\, f^{\prime}(S)g^{\mu\alpha}R_{(\alpha\beta)}g^{\beta\nu}$ is a symmetric
$(2,0)$-rank tensor density of weight $1$ which we additionally assume to be
nondegenerate. This assumption entitles us to introduce a new metric
$h_{\mu\nu}$ by the following definition
\begin{equation}\label{h-met}
\sqrt{\det h}\,h^{\mu \nu }=\sqrt{\det g}\,f^{\prime}(S)g^{\mu\alpha}
R_{(\alpha\beta)}(\Gamma)g^{\beta\nu}
\end{equation}
The metric $h$ is hence  called a Levi-Civita metric since the field equations (\ref{bffv2})  and consequently
$\Gamma$ is the Levi-Civita connection of it: $\Gamma =\Gamma _{LC}(h)$. The Ricci tensor of $h$ can be simply defined as  $R_{(\mu\nu)}(\Gamma)
=R_{\mu \nu }(h)\equiv R_{\mu\nu}$. It should be easily recognized that eq. (\ref{bffv2})
defines $h_{\mu\nu}$ only up to multiplicative constant. Therefore the metric $h$ is not a 
good candidate for a physically meaningful object.\\
The algebraic equation (\ref{bffv1}) can be easily converted into
the matrix form
\begin{equation}\label{P-mat}
P^{2}=\frac{1}{4 f^{\prime }(S)} f(S)I+\frac{\kappa}{2 f^{\prime }(S)} \hat T
\end{equation}
by using the endomorphisms  $P$ and $ \hat{T}$ (i.e. $(1,1)$-tensorfields) as defined before:
\begin{eqnarray}\label{def}
P=P_{\nu }^{\mu }&=&g^{\mu\alpha}R_{(\alpha\nu)}\\
\hat{T}=\hat{T}_{\nu }^{\mu }&=&g^{\mu\alpha}T_{\alpha\nu} \nonumber
\end{eqnarray}
and $I=\delta^\mu_\nu$ denotes the identity endomorphism, i.e. a Kronecker delta in dimension $4$.  In matrix notation
one can also write $P=g^{-1}R$ and $\hat T=g^{-1}T$.\\
Equation (\ref{P-mat}) can be supplemented by the scalar-valued equation
obtained by taking the trace of  (\ref{bffv1}) or of (\ref{P-mat})\footnote{We remark that in this context $S=\mathrm{tr}P^{2}$}
\begin{equation}
f^{\prime }(S)S-f(S)=\frac{\kappa}{2} g^{\alpha \beta }T_{\alpha \beta }\equiv
 \frac{\kappa}{2}\tau
\label{master2}
\end{equation}
which governs solutions of the matrix equations  (\ref{bffv1}) and (\ref{P-mat}) and we will define it as the \textit{structural equation} of spacetime under analysis.
We remark that in the vacuum case, as much as in
the particular case of radiating matter ($\tau=0$), we have that ($\ref{master2}$) gives constant solutions for the values of $S$, so that the universality property of Einstein field equations still holds \cite{BFFV}. In the more general case of interaction of the gravitational field with matter we are considering, we will have that solutions of ($\ref{master2}$) are no longer constants, but they are related with the values of  $\tau$. This means that the solutions of (\ref{master2}) are dependent on the choice of the stress-energy tensor for matter (at least on the trace of the stress-energy tensor) and moreover these solutions are dynamical, since $\tau$ is generally time-dependent. \\
The structural equation ($\ref{master2}$) can be formally (and hopefully explicitly) solved
expressing $S=F(\tau)$. This allows to reinterpret both $f(S)$ and $f' (S)$ as functions of $\tau$ in the  expressions:
\begin{equation} \label{diptau}
\cases{
f (S)=f (F(\tau))=f (\tau)\cr
f^\prime (S)=f^\prime (F(\tau))=f^\prime (\tau)
}
\end{equation}
where, for convenience, we will use in the following the abuse of notation $f (F(\tau))=f (\tau)$ and $f^\prime (F(\tau))=f^\prime (\tau)$.
For any real solution $S=F(\tau)$ of (\ref{master2}) it is hence possible to compute the operator
$P$ by solving the matrix equation (\ref{P-mat}):
\begin{equation}\label{alg2}
P^{2} =\frac{f(\tau)}{4f^{\prime }(\tau)}
 I+\frac{\kappa }{2f^{\prime }(\tau)}\hat T
\end{equation}
by simply calculating a square root of the  endomorphism on the right hand side (under the condition $f^\prime( \tau ) \ne 0$).
The $P$ tensorfield results consequently
to be a function of $P=P(\tau)$, due to (\ref{diptau}).
Owing to the cosmological principle it results that $\tau$ and consequently the operator $P$ will be simply functions of time, once the stress energy tensor for matter is chosen. \\
We remark that the  solution proposed  above for the matrix equation (\ref{alg2}) is just one of the solutions of (\ref{alg2}) and precisely it represents the
 simplest diagonal solution in the set of all possible solutions of (\ref{alg2}). The definition of $P$ given in (\ref{def}) should satisfy some restrictions, deriving directly from field equations of the theory (\ref{bffv1}), as much as $h$ should satisfy them. These conditions can be read in a differential form in  (\ref{def}), which is however unsolvable, or translated into an algebraic expression (\ref{alg2}). This equation thus selects operators which are meaningful in the theory we are constructing. \\
 Off-diagonal solutions can also be found (as much as in \cite{cosmolla}), but in the $4$-dimensional case under
 analysis they are very difficult to be explicitly calculated. For our purposes we restrict thence overslves to the diagonal solution; more
complicated solutions of these equations, in relation with the geometrical structure of spacetime, will be possibly  analyzed in forthcoming papers.  \\
On the other hand equation (\ref{h-met}) tells us the the metric $h$ is conformal
to a symmetric bilinear form; i.e., in matrix notations:
\begin{equation}
h \simeq (g^{-1}Rg^{-1})^{-1}=P^{-1} g
\end{equation}
Now we are in position to calculate the conformal factor, which results to be $\Omega=\frac{\sqrt{ \hbox{det h}}}{\sqrt{\hbox{det g}}} [f^\prime (S)]^{-1}$ and we will have in matrix notations that $h=\Omega P^{-1} g$.  
Owing to the equations (\ref{h-met}) and  (\ref{def}) respectively, it is then  possible
to set:
\begin{eqnarray}\label{detter}
\det R=\det g\cdot\det P\\
f^\prime  (S)^{4}\det R=\det h \nonumber
\end{eqnarray}
If we consider together the above equations (\ref{detter}) it results we see that the conformal factor can be calculated to be $\Omega= f^\prime  (S) \sqrt{ \hbox{det P}}$,
 where  $\det P$ can be simply obtained from the solution of  equation (\ref{alg2}), once the structural equations (\ref{master2}) are solved. At this point we stress again that the conformal factor $\Omega$ is defined only up to an irrelevant multiplicative constant which has no influence on the physically measurable quantities $g$ and $\Gamma$.\\
We are thus able to express the metric $h$ in terms of the operator  $P$ and the physical metric $g$ from (\ref{h-met}),  as:
\begin{equation}\label{h-met2}
h_{\mu \nu }=h_{\mu \nu }(\tau)=f^{\prime }(\tau)\sqrt{\det P(\tau)} \; g_{\mu \alpha }\left( P^{-1}\right)
_{\nu }^{\alpha }
\end{equation}
where we have stressed the dependence of $h$ from $\tau$, which follows from (\ref{master2}).
Once again to obtain this expression for $h_{\mu \nu }$ explicitly we should require  that (\ref{diptau}) can be solved analytically.
Having finally  calculated $P_{\nu }^{\mu }$ and $\det P$ from the algebraic
equations (\ref{alg2}) and (\ref{detter}) (for a given $g_{\mu \nu }$ and $T_{\mu\nu}$)
the generalized Einstein equation ensuing from (\ref{bffv1}) take  the simple form:
\begin{equation}\label{genEin2}
R_{\mu \nu }\left( h\right)=P_{\nu }^{\alpha } g_{\mu \alpha }
\end{equation}
with $h_{\mu\nu}$ given by  (\ref{h-met2}) and now
the physical metric $g$ does not need to be Einstein. This expression for the generalized Einstein equations is formally the same obtained for
non-linear Lagrangians in the generalized  Ricci scalar in (\ref{gen_Ein}). Differences arise in the definition of the operator $P$ (compare expressions (\ref{alg1}) and (\ref{alg2}))
and  the metric $h$  (compare expressions (\ref{h_met1}) and (\ref{h-met2})), which in this last case results, in general, to be no longer conformal to $g$.
For the same reasoning as before one should easily realize that for the matter-free case equation (\ref{genEin2})
becomes just Einstein equation for the metric $h$, with a cosmological constant depending on the analytical form of $f(S)$. We remark once again that in the vacuum case we have that $P$ is proportional to the identity and solutions of (\ref{master2}) are constants.  In the case of interaction with matter both $P$ and $f(S)=f(\tau)$ depend on the stress-energy tensor of matter, i.e. they are both dynamical.  We thus skip from a \emph{static} model equivalent to a standard Einstein theory with cosmological constant to a more complicated  \emph{dynamical}  model, which is no longer analogous to Einstein Gravity.


\section{FRW Cosmology in Ricci squared Gravity \label{FRWGframe}}
\noindent For cosmological applications (as already explained in section \ref{cossec}) one has first to choose the physical metric, which is assumed to be $g$ for the moment, to be the Friedmann-Robertson-Walker (FRW) metric, which (in spherical coordinates) takes the standard form (\ref{RW1R}), i.e.:
\begin{equation}
g=-d t^2+a^2 (t) \Big[ {1 \over {1-K r^2}} d r^2+ r^2 \Big( d \theta^2 +\sin^2 (\theta) d \varphi^2  \Big) \Big]. \label{RW1}
\end{equation}
Another main ingredient of the cosmological model is to choose  the perfect fluid stress-energy tensor for matter, introduced in (\ref{Tmunuuu}) and in a co-moving frame in (\ref{TmunuR}). From the conservation law of the energy-momentum  $\nabla^\mu T_{\mu \nu}=0$ the consequent continuity equation takes the form:
\begin{equation} \label{pro}
\dot{\rho}+3H(\rho+p)=0
\end{equation}
where $H=\frac{\dot a}{a}$ is the \textit{Hubble constant}.
The above continuity equation imposes  standard relations between the pressure $p$, the matter density $\rho$
and the expansion factor $a(t)$ \cite{weinberg}, namely:
\begin{equation} \label{rhopt}
p=w\rho \quad , \quad
\rho=\eta a^{-3(1+w)}
\end{equation}
with a positive constant $\eta>0$. As it is well-known the particular values of the parameter $w\in \{-1,0,{1\over 3}\}$ will correspond to the
vacuum,  dust or radiation   dominated universes. Exotic matters, which are up to now under investigation as possible models for dark energy, admit instead values of $w<-1$ which are supported actually by experimental data \cite{Spergel}. We remark that the above expressions (\ref{rhopt}) imply that both $\rho$ and $p$ depend just on time, while they do not depend on space coordinates as an immediate consequence of the cosmological principle. This implies that
the variable $\tau$ is an implicit function of the cosmic time $t$. In order to find its explicit dependence of time one has to solve the Friedmann equation.\\
From (\ref{RW1}) and (\ref{TmunuR}) it follows that $\hat T$ results to be, using the definition (\ref{def}):
\begin{equation}
\hat{T}_{\nu}^\mu=
\left(
\begin{array}{clcr}
-\rho&0&0&0\\
0&p&0&0\\
0&0&p&0\\
0&0&0&p
\end{array}
\right) \label{Tmunuten}
\end{equation}
All diagonal solutions of (\ref{alg2}) can be thus calculated, using expressions (\ref{RW1}) and (\ref{Tmunuten}):
\begin{equation} \label{Pipi}
P_\mu^\nu = \frac{1}{2}\sqrt{\frac{f(\tau)+2\kappa p}{f^{\prime}(\tau)}}\mathrm{Diag} \left( \epsilon
_{0}\sqrt{\frac{f\left( \tau \right) -2\kappa \rho}{f(\tau)+2\kappa p}},\epsilon
_{1},\epsilon _{2},\epsilon _{3} \right)
\end{equation}
where we have formally expressed $S=F(\tau)$  from (\ref{diptau}), where $\tau=3p-\rho$. We  introduce moreover  $\epsilon_\mu=\pm 1; \;\; \mu=0, \dots ,3$, ensuing from the square root of the operator $P^2$.
Notice that  all possible choices of $\epsilon_\mu$
give rise to all possible diagonal solutions of the matrix equation but still
corresponding to the same solution $S=F(\tau)$. This exhibits a phenomenon
of signature change in $f(S)$ theories (see below and \cite{cosmolla}). Reality condition forces us  to
assume that all three terms
\begin{equation} \label{Pipib}
f^{\prime}(\tau)\neq 0,\ \ \ \ f(\tau) -2\kappa \rho\neq 0\ \ \ \  {\rm and}\ \ \ \ f(\tau)+2\kappa p\neq 0
\end{equation}
have to have at the same time the same (negative or positive) sign. In what following we denote
$\varepsilon=sign\,f^{\prime}(\tau)=sign\,(f(\tau) -2\kappa \rho)=
sign\,(f(\tau)+2\kappa p)$.\\
It is hence  possible to calculate the Levi-Civita metric $h$, which from (\ref{h-met}) turns out to be:
\begin{eqnarray}
h_{\mu \nu} &=& \frac{1}{2}\varepsilon \sqrt{\epsilon |f^{\prime}(\tau)|  [(f(\tau)+2\kappa p)
(f(\tau)-2\kappa \rho)]^{\frac{1}{2}} } \times \\
&\times& \mathrm{Diag} \left(- \epsilon
_{0}\sqrt{\frac{f\left( \tau \right) +2\kappa p}{f(\tau)-2\kappa \rho}}, \frac{\epsilon
_{1} a^2}{1-K r^2},\epsilon _{2} r^2 a^2,\epsilon _{3} r^2 a^2 \sin^2(\theta) \right)\nonumber
\end{eqnarray}
where we denoted  $\epsilon=\epsilon_{0}\epsilon_{1} \epsilon_{2} \epsilon_{3}$.  Neglecting an irrelevent multiplicative constant factor (which can be in general complex or imaginary) the above expression can be suitably rewritten, for convenience, as:
\begin{equation} \label{acca}
h_{\mu \nu} = b(\tau)
\mathrm{Diag} \left(- \epsilon
_{0} c(\tau), \frac{\epsilon
_{1} a^2}{1-K r^2},\epsilon _{2} r^2 a^2,\epsilon _{3} r^2 a^2 \sin^2(\theta) \right)
\end{equation}
where:
\begin{equation} \label{accadef}
\cases{
b(\tau)= \sqrt{ |f^{\prime}(\tau)|}  [(f(\tau)+2\kappa p)(f(\tau)-2\kappa \rho)]^{\frac{1}{4}}  \cr
c(\tau)=\sqrt{\frac{f\left( \tau \right) +2\kappa p}{f(\tau)-2\kappa \rho}}
}
\end{equation}
and $b(t)$ results to be  a \textit{generalized conformal factor}\footnote{It is evident from the above 
expression that the two metrics $h$ and $g$ are no more conformal as they were in the case of the $f(R)$ Lagrangians,
apart from the very particular case of $c(t)=\hbox{const}$. However a suitable redefinition of the cosmic time 
variable restores the conformal relation  between $h$ and $g$} between the two metrics $g$ and $h$, while $c(t)$
describes a  \textit{rescaling factor} for the cosmological time $t$. We notice that both the generalized 
conformal factor $b(t)$ and the rescaling  factor $c(t)$ are positive definite by definition. The change of 
signature is related with 
coefficients $\epsilon=\pm 1$
and the freedom in their choice produces a multiplying of the Friedmann-Robertson-Walker manifold, which could be  related with quantum cosmology phenomena. \\
From the above expression (\ref{acca}) it is possible to notice that some choices of the value of 
$\epsilon_\mu$, which is up to now completely free, will change the signature of the metric $h$, so 
that a signature change process appears as much as in \cite{cosmolla}. If we choose all 
$\epsilon_\mu=\pm 1 $ to be equal we will obtain again a Lorentzian metric, at most with a different 
convention in signs. If any other choice is performed we will possibly have different  signatures 
for the metric (corresponding to Euclidean, Lorentzian or Kleinian signaturs) and the $t$ coordinate may 
then loose its preferred physical significance. \\
The Ricci tensor of the metric $h$ can be calculated from the expression (\ref{acca}); it results to be 
diagonal with the following components:
\begin{eqnarray} \label{ricciacca}
R_{00}&=&{3 \over 4} \left[- 2 {\dot{a} \over a}  {\dot{b} \over b} +2 {\dot{a} \over a}  {\dot{c} \over c}+  
{\dot{b} \over b}  {\dot{c} \over c}+2  \left(  {\dot{b} \over b} \right)^2-2    {\ddot{b} \over b}-4{\ddot{a} \over a}\right] \\
R_{11}&=& {{\epsilon_0 \epsilon_1} \over {c }} {{a^2} \over {4 (1-Kr^2)}} \left[  \left( 10 {\dot{a} \over a} 
 {\dot{b} \over b} -2 {\dot{a} \over a}  {\dot{c} \over c}-  {\dot{b} \over b}  {\dot{c} \over c}+8  \left(  {\dot{a} \over a} \right)^2+2    {\ddot{b} \over b}+4{\ddot{a} \over a}  \right)  +{{8 K \epsilon_0 \epsilon_1 c} \over a^2}\right] \nonumber \\
R_{22}&=&\epsilon_2 {{a^2 r^2} \over {4 }} \left[ {\epsilon_0 \over {c }}  \left( 10 {\dot{a} \over a}  {\dot{b} \over b} -2 {\dot{a} \over a}  {\dot{c} \over c}-  {\dot{b} \over b}  {\dot{c} \over c}+8  \left(  {\dot{a} \over a} \right)^2+2    {\ddot{b} \over b}+4{\ddot{a} \over a}  \right)  +   { \epsilon_1} {{8 K} \over a^2}\right]+{{\epsilon_1-\epsilon_2} \over { \epsilon_1}} \nonumber\\
R_{33}&=&\epsilon_3  {{a^2 r^2 \sin^2 (\theta)} \over {4 }} \left[ {\epsilon_0 \over {c }}  \left( 10 {\dot{a} \over a}  {\dot{b} \over b} -2 {\dot{a} \over a}  {\dot{c} \over c}-  {\dot{b} \over b}  {\dot{c} \over c}+8  \left(  {\dot{a} \over a} \right)^2+2    {\ddot{b} \over b}+4{\ddot{a} \over a}  \right)  +{\epsilon_1} {{8 K} \over a^2}\right]+ \sin^2 (\theta) \epsilon_3 {{\epsilon_1-\epsilon_2} \over { \epsilon_1 \epsilon_2}} \nonumber
\end{eqnarray}
The r.h.s. of the generalized Einstein equations  (\ref{genEin2}) is obtained from (\ref{Pipi}):
\begin{equation} \label{rhsgenei}
P_\mu^\alpha g_{\alpha \nu} = \frac{1}{2}\sqrt{\frac{f(\tau)+2\kappa p}{f^{\prime}(\tau)}}\mathrm{Diag} \left( -\epsilon
_{0}\sqrt{\frac{f\left( \tau \right) -2\kappa \rho}{f(\tau)+2\kappa p}},\frac{\epsilon
_{1} a^2}{1-K r^2},\epsilon _{2} a^2 r^2,\epsilon _{3}a^2 r^2 \sin^2(\theta) \right)
\end{equation}
Comparing expressions (\ref{ricciacca}) and (\ref{rhsgenei}) we obtain that we must impose that $\epsilon _{1}=\epsilon _{2}$, which derives from simple algebraic consistent conditions on the generalized Einstein equations (\ref{genEin2}). It follows that, like in the standard cosmological models, we have only two relevant field equations, corresponding to the $00$ and the $ii$ components. The values of $\epsilon _{0}$ and $\epsilon _{3}$ are completely arbitrary. We remark however the the choice of the value of $\epsilon _{3}$ does not affect field equations as it cancels from field equations, as we will see later. Field equations are fixed once we have chosen the values of $\epsilon _{0}$ and $\epsilon _{1}$ which modify respectively the $00$ and the $ii$ component of field equations. \\
To obtain modified Friedmann equations, we have to  take into account the  relevant generalized Einstein equations, which are for the $00$ component of  (\ref{genEin2}):
\begin{equation} \label{00comfr}
\left[ 2 {\dot{a} \over a}  {\dot{b} \over b} -2 {\dot{a} \over a}  {\dot{c} \over c}-  {\dot{b} \over b}  {\dot{c} \over c}-2  \left(  {\dot{b} \over b} \right)^2+2    {\ddot{b} \over b}+4{\ddot{a} \over a}\right]= \frac{2 \epsilon
_{0}}{3}\sqrt{\frac{f\left( \tau \right) -2\kappa \rho}{f^{\prime}(\tau)}}
\end{equation}
and for the generic the $ii$ component\footnote{We stress again that with the assumptions $\epsilon _{1}=\epsilon _{2}$ each $ii$ component provides the same field equation, as it should be expected.}:
\begin{equation} \label{11comfr}
\left[  \left( 10 {\dot{a} \over a}  {\dot{b} \over b} -2 {\dot{a} \over a}  {\dot{c} \over c}-  {\dot{b} \over b}  {\dot{c} \over c}+8  \left(  {\dot{a} \over a} \right)^2+2    {\ddot{b} \over b}+4{\ddot{a} \over a}  \right)  +\frac{\epsilon_0}{\epsilon_1} {{8 K {c }} \over { a^2}}\right]={2} {c \epsilon_0} \sqrt{\frac{f(\tau)+2\kappa p}{f^{\prime}(\tau)}}
\end{equation}
Subtracting the first equation (\ref{00comfr})  from the second equation (\ref{11comfr}), we obtain that the second derivatives of the scale factor $a$ and of the conformal factor $b$ both disappear and we get the modified Friedmann equations in the form:
\begin{equation} \label{MFE}
\left[      {\dot{a} \over a} +     {\dot{b} \over {2 b}}  \right]^2  =\frac {c \epsilon_0}{4} \sqrt{\frac{f(\tau)+2\kappa p}{f^{\prime}(\tau)}} - \frac{ \epsilon
_{0}}{12}\sqrt{\frac{f\left( \tau \right) -2\kappa \rho}{f^{\prime}(\tau)}}  - {\epsilon_0}{\epsilon_1} {{ K {c }} \over { a^2}}
\end{equation}
where   the expression on the l.h.s. can be defined as a modified Hubble constant (which is moreover analog to (\ref{MHC}); see also  \cite{ABFR}), which rules the dynamical evolution of the universe:
\begin{equation} \label{MFH}
\hat{H}^2=\left[      {\dot{a} \over a} +     {\dot{b} \over {2 b}}  \right]^2
\end{equation}
The r.h.s. of the modified Friedmann equations for Ricci squared theories differs however from the r.h.s. of (\ref{MHC}) for Ricci scalar theories, as it should be reasonably expected. The evolution of the model  is just dependent on the evolution of the scale factor $a(t)$ and of the modified conformal factor $b(t)$ (i.e. on $\hat{H}$), while  derivatives of the factor $c(t)$ disappear. This fact is strongly analogous with the case of Ricci scalar theories.  \\
We remark that, as already observed before, the expression (\ref{MFE}) and field equations depend only on the values of $\epsilon _{0}$ and $\epsilon _{1}$. The sign factor  $\epsilon _{0}$ appears as a constant in front of the r.h.s. of modified Friedmann equations, which can be rewritten as
\begin{equation}
\hat{H}^2=\epsilon _{0}  \left[ \frac {f(\tau)+\kappa \tau+ 2 \kappa \rho}{6 \sqrt{f^\prime (\tau) [f(\tau)-2 \kappa \rho]}}  - {\epsilon_1} {{ K {c }} \over { a^2}} \right]
\end{equation}
so that a suitable choice of $\epsilon _{0}$ allows the r.h.s. to be always positive as expected. 
In fact $\epsilon_0$ has to be chosen in accordance with the prescription 
$\epsilon_0= sign\,f^{\prime}(\tau) (f(\tau)+\kappa \tau+ 2 \kappa \rho)$, provided  the conditions 
(\ref{Pipib}) are satisfied. However, we see that on the other hand $\epsilon _{1}$ appears only in the 
term related with the curvature $K$ of the spacelike hypersurface. As it is obvious also from the 
explicit expression of $h$ (\ref{acca}), choosing different values of  $\epsilon _{1}$ is equivalent 
to change the sign of the spatial curvature.  We remark finally that the choice of $\epsilon_3$ is 
irrelevant for field equations.


\section{Polynomial Lagrangians in the Ricci squared invariant}
\begin{figure}[tbp]
\centering
\includegraphics[scale=0.65]{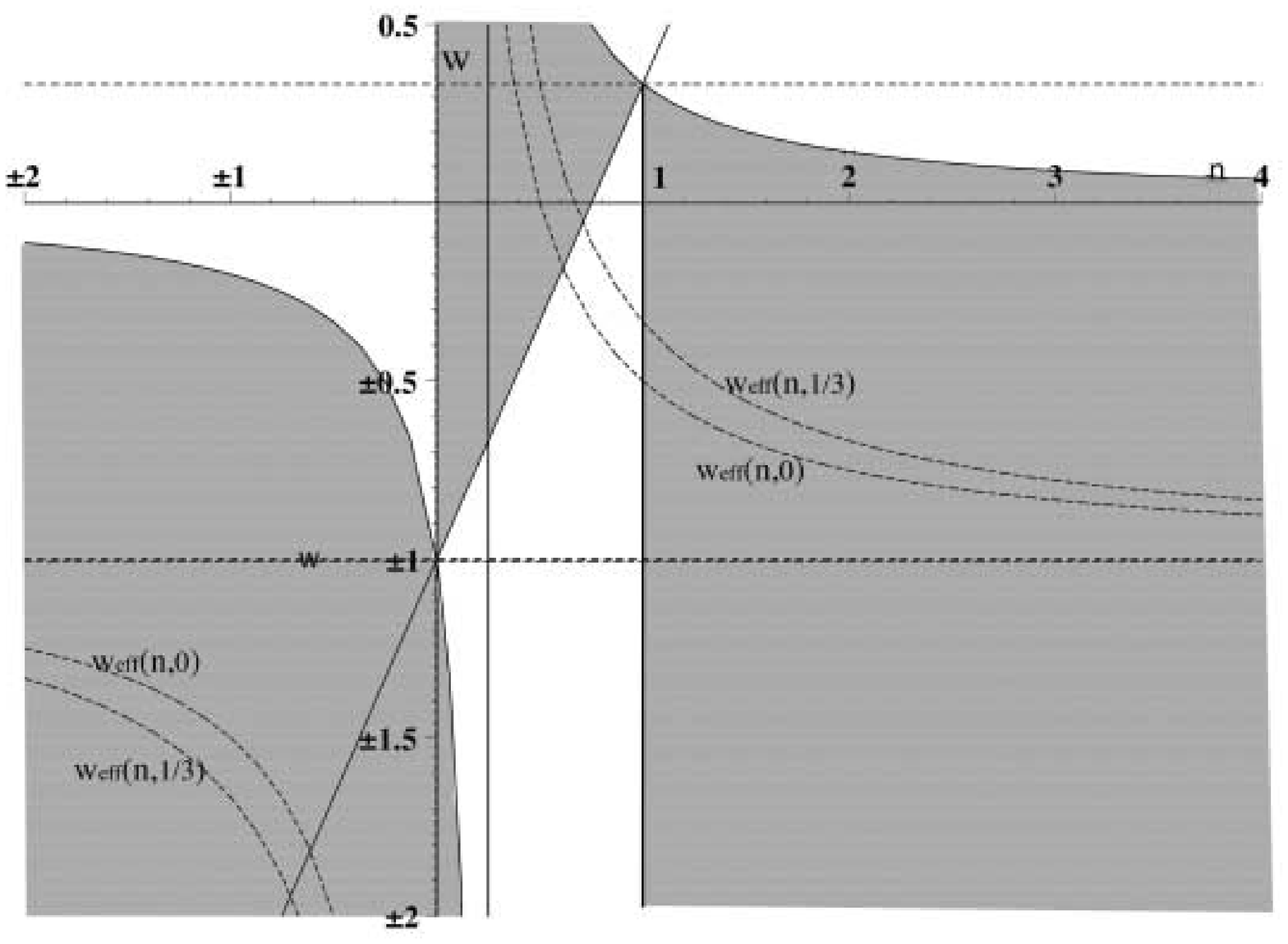}
\caption{\label{diagr} \fs Phase portrait for the plane $(w, n)$, where $w$ is the fluid parameter characterizing matter and $n$ the exponent of $S$ in the Lagrangian. The shadow areas  represent physically and mathematically admissible pairs. We notice the following:\\
- for $n>1$ we cannot have radiation ($w=\frac{1}{3}$) but dust  is allowed;\\
-for dust matter ($w=0$) and $n>1$, $w_{eff} \rightarrow -1^+$, i.e., $w_{eff}$ can be approach only from above;\\
- {in the contrary negative powers ($n<0$) do not allow dust, although any dust-like
	matter can be allowed for large enough $|n|$. Moreover in this case $w_{eff} \rightarrow -1^-$ is possible	from below (super-acceleration).}}
 \end{figure}

We choose, as a relevant example to deal with, polynomial Lagrangians in $S$. In strict analogy with what has already been done for the Ricci scalar theories,
polynomial Lagrangians can be considered as approximations\footnote{We are particularly interested in the 
cases of very small and large values of $S$, reproducing the cases of large and small curvatures of the 
universe, owing to the (linear) quadratic relation between $S$ and $R_{\mu \nu}$.} of any analytical 
expression in $S$ in the suitable limit \cite{ABFR}. It is hence worth investigating the
behavior of cosmological solutions of Ricci squared theories described by means of Lagrangians 
which are pure powers of $S$:
\begin{equation} \label{pupow}
f(S)=\beta S^n
\end{equation}
As a matter of facts polynomial Lagrangians can be approximated to pure power Lagrangians if the asymptotical behavior is considered
and just the first leading term is taken into account.  \\
From the structural equations (\ref{master2}) we  obtain that for the above pure power Lagrangian (\ref{pupow}) in the Ricci squared invariant:
\begin{equation}
\cases{
f(\tau)=\frac{\kappa}{2  (n-1)} \tau \cr
f^\prime (\tau)=\frac{n\kappa\tau}{2\,(n-1)}\left[\frac{\kappa \tau}{2 \beta (n-1)}\right]^{-\frac{1}{n}}}
\end{equation}
where we have to require $n \ne 1$ to avoid singularities in the theory (this implies that the case $f(S) \simeq S$ is not allowed\footnote{This is similar to the presence of a methodological singularity for $f(R) \simeq R^2$ in non-linear theories of Gravity depending on the Ricci scalar; see e.g.  \cite{staro} and \cite{ABFR}}). Since in the physically interesting cases one has $\tau\leq 0$ we see that  for generic $n$ we have to assume
$$\beta(n-1)<0
$$
However, for odd $n$ we can allow $\beta(n-1)>0$ (see also \cite{ABFR} in this context). Taking into account the standard relations:
\begin{equation} \label{rhopt1}
\tau=(3w-1) \rho \quad , \quad p=w\rho \quad , \quad
\rho=\eta a^{-3(1+w)}
\end{equation}
and performing straightforward calculations we obtain from (\ref{accadef})
the generalized conformal factor (up to multiplicative constant):
\begin{equation}
b(t) \simeq a^{-3(1+w)(1-\frac{1}{2n})}
 \end{equation}
Performing further calculations by means of (\ref{MFH}) it is simple to obtain:
 \begin{equation} \label{hRenne}
\hat{H}^2= \left[ \frac{(3w+1)(2n-1)-2}{4 n} \right]^2 H^2
 \end{equation}
 We remark that in the particular case of $w=-1$ the expression (\ref{hRenne}) implies that $\hat{H}^2=  H^2 $, 
 independently on the value of $n$.  Using the same relations, the r.h.s. of equation (\ref{MFE}) results to be:
 \begin{equation} \label{lhopt}
\hat{H}^2= \frac{\epsilon_0}{6} \frac{(3w+1)(2n-1)-2}{ \sqrt{n(3w-1) (3w+3-4 n)}}  \left[ \frac{\kappa  (3w-1)}{2 \beta (n-1)}  \right]^{\frac{1}{2n}} \rho^\frac{1}{2n}- \frac{\epsilon_0}{\epsilon_1} \frac{K}{a^2} \sqrt{ \frac{4nw-(w+1)}{3(1+w)-4 n}}
 \end{equation}
We stress that the rescaling factor $c(t)$ is, in this particular example, independent on time:
 \begin{equation}
c(t)=\sqrt{\frac{4wn-w-1}{3w+3-4n}}
 \end{equation}
Combining  equations (\ref{hRenne}) and  (\ref{lhopt}) together we obtain that the Hubble ${H}^2$ 
constant for the physical metric $g$ is:
\begin{equation} \label{H2pow}
{H}^2=\left(    {\dot{a} \over a}   \right)^2 = \epsilon_0 P(n,w) 
a^{-\frac{3(1+w)}{2n}}- \frac{\epsilon_0}{\epsilon_1} Q(n,w){K}a^{-2}
\end{equation}
where we have defined:
\begin{equation}
\cases{
P(n,w)= \frac{ 8 n^2}{3 \sqrt{n(3w-1) (3w+3-4 n)}[(3w+1)(2n-1)-2]}  \left[ \frac{\kappa  (3w-1)}{2 \beta (n-1)} \eta \right]^{\frac{1}{2n}}  \cr
Q(n,w)=\sqrt{ \frac{4nw-(w+1)}{3(1+w)-4 n}}  \left[ \frac{4 n}{(3w+1)(2n-1)-2}\right]^{2}
}
\end{equation}
From the above expression the deceleration parameter can be calculated by means of the standard formula already introduced in (\ref{qhub}) and it can be formally calculated from (\ref{H2pow}) under the form\footnote{We say that the deceleration parameter can be formally calculated, as we do not know a priori if any physical solution exists in all cases considered.}:
\begin{equation} \label{decpow}
q(t,w,n) =-1+ \frac{\frac{3(1+w)}{4n}\epsilon_1 P(n,w) 
a^{-\frac{3(1+w)}{2n}}- Q(n,w){K}a^{-2} }{ \epsilon_1 P(n,w) \eta^\frac{1}{2n} a^{-\frac{3(1+w)}{2n}}-  Q(n,w){K}a^{-2}}
\end{equation}
We obtain consequently that when the $a^{-2}$ term dominates over $a^{-\frac{3(1+w)}{2n}}$ the deceleration parameter results to be positive, i.e. $q(t,w,n)\rightarrow 0^+$, while when the term $a^{-\frac{3(1+w)}{2n}}$ dominates over $a^{-2}$ or in the physically very important case $K=0$, the deceleration parameter will be:
\begin{equation} \label{decpowlim}
q(w,n) =-1+ \frac{3(1+w)}{4n}
\end{equation}
This implies that $q(w,n)$ is negative for $n<0$ or $n>\frac{3(1+w)}{4}>0$, owing to the positivity of 
$(1+w)>0$ for standard matter. Taking also into account restriction on paprameters $(w, n)$ coming from 
(\ref{Pipib}) the whole situation can be visualized on the phase diagram FIG. \ref{diagr}.

Comparing expression (\ref{H2pow}) and the standard relation which derives from General Relativity (see  \cite{carrol2} and \cite{ABFR}) it is easy to obtain that the effective value of $w_{eff}$ deriving from Ricci squared alternative theories of gravity is:
\begin{equation} \label{weffpow}
w_{eff}(w,n) =-1+ \frac{(1+w)}{2n}
\end{equation}
Both  limiting values for $w_{eff}(\frac{1}{3},n)$ and $w_{eff}(0,n)$ are marked on the phase diagram FIG. \ref{diagr}.\\

We can compare the values of $q(w,n)$ and $w_{eff}(w,n)$ obtained in this case with the values of $q(w',n')$ (\ref{qqqquti}) and $w_{eff}(w',n')$ (\ref{weffR}) obtained for alternative theories of Gravity with pure power Lagrangians  of the Ricci scalar $F(R)=\beta' R^{n'}$, already treated in \cite{ABFR}. It turns out that they differ just for a  factor if the value of $w$ is assumed to be fixed, which is equivalent to state that we are dealing with the same kind of matter. It is simple to see that:
\begin{equation}
\cases{
q(w,n)=q(w',n')\cr
w_{eff}(w,n)=w_{eff}(w',n')
}
\Longleftrightarrow
n={n' \over 2}
\end{equation}
These results generalize and confirm  the results already obtained in \cite{miic} in the very particular case of quadratic Lagrangians.


\subsection{Polynomial Lagrangians}

\noindent As we already stated before, pure-powers Lagrangians  in the Ricci squared invariant can
be considered as \textit{approximations} of more physical polynomial-like Lagrangians of the type:
\begin{equation}
f(S)=S+\frac{\alpha}{(1+n)S^n}+\frac{\beta}{1-m} S^m
\end{equation}
(here both $n>0$ and $m>0$, with $m \ne 1$.\\
We just consider for simplicity the case of  flat universe $K=0$. In the limit of small or large curvatures,
corresponding to the cases of present time universe and early time universe,
 we obtain from the structural equations  that the leading terms are respectively:
$$
\cases{
S \rightarrow 0 \Rightarrow \frac{-\alpha}{S^n}=\kappa \tau\cr
S \rightarrow \infty \Rightarrow  -\beta S^m=\kappa \tau}
$$
From (\ref{decpowlim}) we deduce that polynomial Lagrangians provide an
explanation for early time inflation assuming that $m> {3\over 4}$
and they can provide an explanation to present time cosmic
acceleration assuming that  some inverse power of the generalized Ricci squared curvature invariant
is also present in the Lagrangian (i.e. $\alpha \ne 0$). \\
This result generalizes previous results which have been obtained for Ricci scalar alternative theories of gravity \cite{metricfR}, \cite{ABFR}, \cite{palatinifR} and they are related to the so-called Starobinsky inflation \cite{staro}.


\section{Changing frame}
We have developed up to now a first order \'a la Palatini theory which after appropriate reduction turns out to be based, as we remarked before, on a bi-metric  spacetime with an almost complex structure. In Section \ref{FRWGframe}
we have assumed $g$ to be the "physical" FRW metric. However we do not know a priori which is the most appropriate frame in the bi-metric structured spacetime we have constructed in Section \ref{R2bemetr}. This is the same problem already studied and examined in \cite{ABFR},  \cite{Flanagan} and \cite{Magnano} in the case of Lagrangians depending on the Ricci scalar, where different frames result to be somehow inequivalent.  We shall not comment here on this equivalence problem and refer the reader to a recent interesting discussion by Flanagan (\cite{Flanagan2} and ref.s quoted therein). In our understanding this important problem should be analyzed in more detail, also in relation with the  physical consistency requirements and the presence of instabilities; we plan to treat it in a forthcoming paper \cite{ABFM2}.\\
In this framework it is thus worth considering also the case when the "Jordan" $h$ is chosen to be the physical FRW metric, as we have already done in \cite{ABFR}. More precisely, according to (\ref{acca}) we set:
\begin{equation}
h=  -\epsilon_0d \tilde{t}^2+A^2 (\tilde t) \Big[ {\epsilon_1 \over {1-K r^2}} d r^2+ r^2 \Big( \epsilon_1d \theta^2 +\epsilon_3\sin^2 (\theta) d \varphi^2  \Big) \Big]
\end{equation}
to be, modulo signature, FRW metric 
with a new cosmic time $ \tilde t$ and a new scale factor $A$. The generalized Einstein equation (\ref{gen_Ein}) can be also calculated in
$({\tilde t}, x^i)$- coordinates. This is equivalent to  the assumption
that the metric $h$ is the physical one (i.e., that we can use conformal Jordan frame instead of the original Einstein frame).  In this case, one has to restore the standard Lorentzian signature by setting $\epsilon_0=\epsilon_i=\pm 1$. We consequently obtain:
\begin{equation}
3\frac{\ddot A}{A}=
\frac{\epsilon_0}{c(\tau)b(\tau)}\left (\frac{f\left ( \tau \right) -2\kappa \rho}{4f^\prime(\tau)}\right )^{\frac{1}{2}}
=\frac{\epsilon_0}{2|f^\prime(\tau)|}\left (\frac{f\left ( \tau \right) -2\kappa \rho}{f(\tau)+2\kappa p}\right )^{\frac{3}{4}}
=\frac{\epsilon_0}{2|f^\prime(\tau)|} c(\tau)^{-\frac{3}{2}}
\end{equation}
for the $00$ component while for the $11$ component we find
\begin{equation}
\frac{\ddot A}{A}+2\Big(\frac{\dot A}{A}\Big)^2 +2\frac{K}{A^2}=
\frac{\epsilon_0}{b(\tau)}\left (\frac{f\left ( \tau \right) +2\kappa p}{4f^\prime(\tau)}\right )^{\frac{1}{2}}
=\frac{\epsilon_0}{2|f^\prime(\tau)|}\left (\frac{f\left ( \tau \right) +2\kappa p}{f(\tau)-2\kappa \rho}\right )^{\frac{1}{4}}
=\frac{\epsilon_0}{2|f^\prime(\tau)|} c(\tau)^{\frac{1}{2}}
\end{equation}
where $\dot A$ denotes now the differentiation with respect to the new
cosmic time $\tilde{t}$. We have also taken into account that in this case $d\tilde{\tau}^2=c(\tau)b(\tau)dt^2$
and $A^2=b(\tau) a^2$.

Now the analogue of the Friedmann equation  takes the form
\begin{eqnarray}
\tilde{H}^2 =
\frac{\epsilon_0c^{\frac{1}{2}}}{12|f^\prime(\tau)|}\left (3-c^{-2}\right )
  - \frac{K}{A^2}
\label{hubHfram}
\end{eqnarray}
with $\tilde{H}=\frac{\dot A}{A}$ being the Hubble constant of the conformal metric $h$. 
Thus up to now arbitrary sign factor $\epsilon_0=\pm 1$ can be adjusted as $\epsilon_0=sign\,(3-c^{-2})$ in order to preserve the positivity. \\
We specialize now to the case of pure power Lagrangians in the Ricci squared curvature invariant and in the meanwhile, as already stated before,
to the case of polynomial Lagrangians in some suitable limit. Choosing, as already done in (\ref{pupow}), the Lagrangian to be:
\begin{equation}
f(S)=\beta S^n
\end{equation}
we obtain respectively $ f(\tau)=\frac{\kappa}{2  (n-1)} \tau $ and $ f^\prime (\tau)=n \beta \left[ \frac{\kappa \tau}{2 \beta (n-1)}\right]^{\frac{n-1}{n}} $.
 It follows from equation (\ref{hubHfram}) that the modified Friedmann equation in this frame is:
\begin{equation}
\tilde{H}^2=\tilde P(w,n) \; A^\lambda - \frac{K}{A^2}
\end{equation}
where for convenience sake we have defined the coefficient $\tilde P(\eta, w,n) $  as:
\begin{equation}
\tilde P(w,n) = \frac{|2 \; n(1+3 w)-3(w+1)|}{6\beta\;|n| \; 
|4 w n- w -1|^\frac{3}{4} |3 w +3 -4 n|^\frac{1}{4}  }
\left[ \frac{2 \beta (n-1)}{\kappa\eta(3w-1)} \right]^\frac{n-1}{n}
\end{equation}
and the exponent $\lambda$ as
\begin{equation}
\lambda=\frac{12(w+1)(n-1)}{3(w+1)-2n(3w+1)}
\end{equation}
We can consequently obtain the deceleration parameter by means of formula (\ref{decpow}):
\begin{equation}
\tilde{q} (\tilde t)=- \frac{(1+\frac{\lambda}{2}) \tilde P A^\lambda}
{{\tilde P}  A^\lambda - KA^{-2} }
\end{equation}
This implies that, in the limit when the term 
$A^{-2}$ is dominating over $A^\lambda$  we will have $\tilde{q} (t) \rightarrow 0^+$.
Otherwise in the limit when the term 
$A^\lambda$ is dominating over $A^{-2}$ or in the case $K=0$,  we will have 
$$\tilde{q}(\tilde t)\rightarrow
\tilde{q} (w,n) = -\frac{4n-3(w+1)}{4n-3(w+1)(2n-1)}$$
Thus
$$
\tilde w_{eff}(w,n)=\frac{2n(w-1)+w+1}{4n-3(w+1)(2n-1)}
$$
which for big $n$ behaves as
$$
\tilde w_{eff}(w,\pm\infty)=\frac{1-w}{3w+1}
$$


\section{Conclusions and perspectives}
\noindent In this paper we have analyzed alternative theories of gravity depending on a Lagrangian assumed to be  a general function of the generalized Ricci squared curvature invariant $S$ constructed out of a dynamical metric 
$g$ and a dynamical (torsionless) connection $\Gamma$. The Palatini formalism provides first
order field equations for the metric and the connection $\Gamma$. A structural metric $h$ is introduced, so that the connection turns out to be the Levi-Civita connection of $h$ and $h$ is consequently a Levi-Civita metric. A convenient spacetime bi-metric geometry is thus defined by means of generalized Einstein equations and it is controlled  by means of structural equations;   signature changing phenomena appear. This implies that the metric $h$ can be either a Lorentzian, an Euclidean or a Norden metric, giving us an immediate and natural insight into quantum cosmology theories.   \\
To treat explicitly cosmological models we choose $g$ to be a Robertson-Walker metric and the stress-energy tensor to be the stress-energy tensor of a perfect fluid.  This allowed us to obtain modified Friedmann field equations and a modified Hubble constant related to a (generalized) conformal transformation factor $b(t)$ between $g$ and $h$ and to a rescaling factor for the cosmological time $c(t)$. The metric $h$ can be considered  to be FRW, too, so that it can be conveniently considered as a physical metric in place of the original $g$. Generalized Friedmann equations are obtained also in this framework.\\
If we moreover specialize to the pure-power  case  $f(S)=\beta S^n$ (with $n$ an arbitrary real exponent) we have seen that, with suitable choices of the parameters involved, these models are able to explain the current acceleration of the universe. We obtain moreover that polynomial Lagrangians in the generalized Ricci squared invariant provide an explanation for the inflation of the universe in  suitable limits \cite{Nojiri}.  \\

This paper was thus devoted to analyze the geometrical structure of spacetimes described by means of Ricci squared Lagrangians in
interaction with matter; cosmological applications of this models have been analyzed, following the ideas of \cite{staro} and generalizing the
effort to understand current acceleration of the universe in alternative theories of gravity \cite{carrol2} and \cite{ABFR}. The relation between the geometrical bi-metric structure of spacetime (and in particular the signature change phenomena) and its cosmological implications is very rich in mathematical and physical significance and will form the  subject of future investigations.

\section{Acknowledgements}
\noindent We are very grateful to Prof. S.D. Odintsov for useful discussions and very important suggestions, concerning the physical properties of the theory  considered.  We are moreover very grateful to Prof. S. Capozziello and Prof. G. Magnano for helpful remarks. \\
This work is partially supported  by GNFM--INdAM research project ``\emph{Metodi geometrici
in meccanica classica, teoria dei campi e termodinamica}'' and by MIUR: PRIN 2003 on
``\emph{Conservation laws and thermodynamics in continuum mechanics and field theories}''.
Gianluca Allemandi is supported by the I.N.d.A.M. grant: "Assegno di collaborazione ad attivit\'a di ricerca a.a. 2002-2003".


\end{document}